\documentstyle[12pt]{article}
\newcommand{\be}{\begin{equation}}
\newcommand{\ee}{\end{equation}}
\newcommand{\ie}{{\it i.e. }}
\newcommand{\Z}{Z \!\!\! Z}
\begin{document}
\pagestyle{empty} \noindent
\hspace*{115mm} \normalsize YCTP-P13-94\\ \hspace*{115mm} October 1994\\
\begin{center} \vspace*{30mm} \LARGE World-sheet aspects of mirror
symmetry\footnote[1]{Talk given at the Oskar Klein centenary symposium 19-21
September 1994 in Stockholm, Sweden, to appear in the proceedings.}\\

\vspace*{20mm} \large M\aa ns Henningson\footnote[2]{email:
mans@genesis5.physics.yale.edu}\\
\vspace*{5mm} \normalsize \it Yale University\\ Department of Physics\\ P. O.
Box 208120\\ New Haven, CT 06520-8120 \end{center}

\begin{center}\vspace*{40mm} \large Abstract \\ \end{center}
The first half of this talk is a non-technical discussion of some general
aspects of string theory, in particular the problem of compactification. We
also give an introduction to mirror symmetry. The second half is a brief
account of two recent papers on this subject; one by the author on mirror
symmetry for Kazama-Suzuki models and one by P. Berglund and the author on a
search for possible mirror pairs of Landau-Ginzburg orbifolds.

\newpage \pagestyle{plain} \begin{center}
{\bf WORLD SHEET ASPECTS OF MIRROR SYMMETRY}

\vspace*{10mm}

M{\AA}NS HENNINGSON\\
{\small Yale University, Department of Physics, P. O. Box 208120\\
New Haven, CT 06520-8120, USA}

\vspace*{10mm}
ABSTRACT

\vspace*{5mm}
\parbox{127mm}
{\small

The first half of this talk is a non-technical discussion of some general
aspects of string theory, in particular the problem of compactification. We
also give an introduction to mirror symmetry. The second half is a brief
account of two recent papers on this subject; one by the author on mirror
symmetry for Kazama-Suzuki models and one by P. Berglund and the author on a
search for possible mirror pairs of Landau-Ginzburg orbifolds.}
\end{center}

\vspace*{10mm} \noindent
{\bf 1. String theory, compactification and mirror symmetry}\\

String theory is an attempt to construct a unified theory of all kinds of
matter and all types of interactions, including gravity
\cite{Green-Schwarz-Witten}. It naturally incorporates many attractive ideas in
particle physics, such as higher dimensions, grand unification and
supersymmetry. The most remarkable accomplishment of string theory is that it
seems to provide a consistent way of incorporating Einstein's theory of
gravity, \ie general relativity, in a quantum mechanical framework. The basic
idea of string theory is that the fundamental constituents of nature are
one-dimensional objects, strings, rather than zero-dimensional point particles.
A string sweeps out a two-dimensional world-sheet as it propagates through
space-time. This is the analogue of the one-dimensional world-line of a point
particle.

We can get an estimate of the typical size of a string by simple dimensional
analysis: The characteristic scale of string theory, where we could expect
`stringy' effects to be important, is given by the Planck length
$\sqrt{\frac{\hbar G}{c}} \simeq 10^{-35}$ m. This is about twenty orders of
magnitude shorter than the length scales relevant to ordinary particle physics,
where we expect string theory to be well approximated by some low energy
effective field theory.

When one examines the quantum mechanical consistency of (supersymmetric) string
theory, it turns out that the dimension of space-time must be ten rather than
the observed four. Inspired by the work of Kaluza and Klein
\cite{Kaluza}\cite{Klein} on higher dimensional theories, we resolve this
apparent paradox by postulating that the ground state of string theory takes
the form
\be
{\rm 10\!-\!dimensional \; string \; vacuum} = M_4 \times K,
\ee
where $M_4$ is the ordinary 4-dimensional Minkowski space and $K$ is some
6-dimensional, compact, Planck-sized space.

A closer examination reveals that $K$ must be a so called Calabi-Yau space
\cite{Candelas-Horowitz-Strominger-Witten}. These spaces are of great
mathematical interest because of their remarkable geometric properties. Their
existence was conjectured by Calabi in the 1950's and proved by Yau in 1977
\cite{Yau}. Given a Calabi-Yau space $K$, there are in general two different
types of continuous deformations that preserve the Calabi-Yau conditions; we
will refer to them loosely as `shape' deformations and `size' deformations. The
number of independent such deformations depends on the particular example
studied.

The topology and geometry of the Calabi-Yau space $K$ determine the properties
of the observable low energy physics in Minkowski space, \ie the unbroken
gauge-group, the particle spectrum, the Yukawa couplings etcetera. It is often
convenient to think of this situation as follows: The string propagating on a
Calabi-Yau space defines a 2-dimensional, $N = 2$ superconformal field theory,
which in its turn determines a Minkowski space low energy effective field
theory.

The intermediate step involves a two-dimensional field theory defined on the
world-sheet of the string. This has to be a conformal field theory for the
string interpretation to be consistent. The requirement of $N = 2$
supersymmetry on the world-sheet is almost equivalent to $N = 1$ supersymmetry
in Minkowski space, which is considered to be an attractive feature of the low
energy theory. The picture of a string propagating on a Calabi-Yau space is
intuitively appealing, although, as we will see later, one should not
necessarily attach too much meaning to this space. The hard facts of this
particular compactification are encoded in a more abstract way in the $N = 2$
superconformal field theory, whereas the observational consequences follow from
the low energy effective field theory.

A string propagating on a Calabi-Yau space $K$ thus determines an $N = 2$
superconformal field theory (and also a low energy effective field theory), but
the reverse is not true. In fact, to account for all $N = 2$ superconformal
field theories we will have to generalize the concept of a Calabi-Yau
compactification, as we will se later. Furthermore, we will see that different
Calabi-Yau spaces (in this generalized sense) may give rise to isomorphic $N =
2$ superconformal field theories and thus equivalent low energy physics.

We will now present the canonical example of a Calabi-Yau space: A quintic
hypersurface in ${\bf CP}^4$.  We start by considering complex 5-space ${\bf
C}^5$ with coordinates $X_1, \ldots, X_5$. To get an intuitive idea of what is
going on, it might be helpful to think of an analogy with ${\bf C}^5$ replaced
by ordinary real 3-space ${\bf R}^3$. Complex projective 4-space ${\bf CP}^4$
could be defined as the space of all complex lines through the origin in ${\bf
C}^5$.

Alternatively we may think of it as ${\bf C}^5$  minus the origin modulo the
equivalence relation $(X_1, \ldots, X_5) \sim \lambda (X_1, \ldots, X_5)$ for
any non-zero complex number $\lambda$.  The analogous construction for ${\bf
R}^3$ would be real projective 2-space ${\bf RP}^2$, \ie the space of lines
through the origin in ${\bf R}^3$. Such a line could be represented by its
intersection with the upper unit half sphere, so  ${\bf RP}^2$ is homeomorphic
to this half sphere with antipodal boundary points identified. Going back to
${\bf CP}^4$, let $W(X_1, \ldots, X_5)$ be a degree 5 polynomial, {\it e.g.}
\be
W(X_1, \ldots, X_5) = X_1^5 + \ldots + X_5^5 + \epsilon X_1 \ldots X_5,
\label{Wex}
\ee
where $\epsilon$ is a parameter. Although $W$ does not make sense as a function
on ${\bf CP}^4$, we may talk of its set of zeros. The hypersurface $K$ in
${\bf CP}^4$ defined by $W(X_1, \ldots, X_5) = 0$ obviously has complex
dimension 3, \ie real dimension 6, and turns out to be a Calabi-Yau space. In
our analogy, a hypersurface in ${\bf RP}^2$ is simply a curve on the upper half
sphere.  One should note that the only thing that really matters are the
intrinsic properties of the Calabi-Yau space $K$ and not its embedding in ${\bf
CP}^4$ or the construction of this space from ${\bf C}^5$. Alternatively, we
could have presented $K$ in some other way, {\it e. g.} by an atlas and
transition functions from one coordinate chart to another.

The parameter $\epsilon$ in (\ref{Wex}) is one of a total of 101 `shape'
parameters in this example. The remaining `shape' parameters also amount to
adding some degree 5 terms to the polynomial $W(X_1, \ldots, X_5)$. Calabi's
conjecture (proven by Yau) amounts to the existence of a `Calabi-Yau' metric
with special properties on the space $K$. This metric is unique up to an
overall constant $r$, which turns out to be the only `size' parameter in this
example. It should be noted that the exact form of the Calabi-Yau metric is not
known.

We have already mentioned that the concept of a Calabi-Yau compactification
needs to be generalized to cover all possible string compactifications. This
issue is closely related to the possible values of the `size' parameter $r$, as
we will now explain. The Calabi-Yau interpretation in fact works best when $r$
is much larger than the Planck length. For an $r$ of the order of the Planck
length, we get a strongly coupled sigma model, which still makes sense,
although the theory is harder to solve.

In fact, we may even continue to negative values of $r$. Here, the geometric
interpretation obviously breaks down, but interestingly enough the $N = 2$
superconformal field theory is still well-defined. If we continue to a negative
value of $r$ with magnitude much larger than the Planck length, we eventually
reach a special point in the parameter space called the Landau-Ginzburg point.
We will give an interpretation of the theory at this point shortly.

The above process has some similarity with the theory of phase transitions, and
we may think of going from a `Calabi-Yau phase' to a `Landau-Ginzburg phase' by
changing the parameter $r$ \cite{Witten-1}. One should note, however, that the
singularity at $r = 0$ may be avoided by giving $r$ an imaginary part, so there
need not be any discontinuous phase transition between the two `phases', much
as we may go from the liquid to the steam phase of a fluid in a smooth way.

To understand the $r \rightarrow - \infty$ limit we will first describe what
might be called the $W$ Landau-Ginzburg theory. We may think of a string
propagating in complex 5-space ${\bf C}^5$ with coordinates $X_1, \ldots, X_5$
under the influence of a potential of the form
\be
W(X_1, \ldots, X_5) + {\rm complex  \;\; conjugate}. \label{pot}
\ee
Here $W(X_1, \ldots, X_5)$ is again a degree 5 polynomial such as (\ref{Wex}).
To complete the definition of the theory, we should also choose a kinetic
energy term. Arguments based on the renormalization group show that there
exists a  kinetic energy term such that the resulting theory is an $N = 2$
superconformal field theory, but the exact form of this kinetic energy is in
general not known. This is analogous to the problem with the form of the
Calabi-Yau metric for a Calabi-Yau space.

The potential (\ref{pot}) with $W$ a degree 5 polynomial such as (\ref{Wex}) is
invariant under $\Z_5$ acting as
\be
(X_1, \ldots, X_5) \rightarrow (\alpha X_1, \ldots, \alpha X_5)
\ee
for $\alpha$ a fifth root of unity, \ie $\alpha^5 = 1$. Given a quantum field
theory invariant under some group of symmetries, we may construct a new theory
by taking the orbifold of the original theory with respect to the symmetry
group \cite{Dixon-Harvey-Vafa-Witten}. In a Lagrangian framework, where a
theory is given by a path integral over some fields and an action functional of
the fields, this amounts extending the domain of integration in the path
integral by allowing the fields to be well-defined only modulo the action of
the symmetry group. In our example, we may construct the Landau-Ginzburg
orbifold $W/\Z_5$ by allowing the fields $X_1, \ldots, X_5$ to transform by an
element of $\Z_5$  as we traverse a non-contractible, closed curve on the
world-sheet.

We may now state the Landau-Ginzburg Calabi-Yau correspondence \cite{Witten-1}
for our example:

\vspace*{3mm} \begin{center} \parbox{127mm}

{\it The $W/\Z_5$ Landau-Ginzburg orbifold theory is equivalent to the $r
\rightarrow - \infty$ limit of the Calabi-Yau theory given (for $r > 0$) by the
hypersurface $W = 0$ in ${\bf CP}^4$. }
\end{center}

\vspace*{3mm}

We have already mentioned the fact that two different Calabi-Yau models (or
Landau-Ginzburg generalizations thereof as discussed above) may give rise to
isomorphic $N = 2$ superconformal field theories and thus equivalent low energy
physics. Mirror symmetry is the best known example of this phenomenon
\cite{Yau-ed.}.  It can be made plausible by the following argument
\cite{Dixon}\cite{Lerche-Vafa-Warner}: Consider again the `shape' and `size'
deformations of a Calabi-Yau space. At the level of $N = 2$ superconformal
field theory, these deformations correspond to a perturbation of the theory by
some operator.

The operators in the theory are characterized by a quantum number that we may
call `charge'. It turns out that a `shape'  or a `size' deformation corresponds
to a perturbing operator of charge  $+1$ or $-1$ respectively. (At the level of
the low energy effective field theory, these deformations of course correspond
to a continuous change of  some coupling constants.) It now seems unnatural
that the profound difference between a `shape' and a `size' deformation of a
Calabi-Yau space should correspond to the purely conventional difference
between operators of  charge $+1$ or $-1$. This observation naturally leads us
to postulate the mirror hypothesis:

 \vspace*{3mm} \begin{center} \parbox{127mm}
{\it To every Calabi-Yau space $K$ corresponds a mirror space $\tilde{K}$ such
that the corresponding $N = 2$ superconformal field theories are isomorphic
(with the isomorphism amounting to a change of sign of the charge quantum
number). }

\end{center}

\vspace*{3mm} Next, we note that a `shape' deformation of $K$ is equivalent to
a charge $+1$ perturbation of the corresponding $N= 2$ superconformal field
theory and thus, by the isomorphism, a charge $-1$ perturbation of the mirror
$N = 2$ superconformal field theory, which corresponds to a `size' deformation
of $\tilde{K}$. The deformed spaces are still mirror partners, and we may thus
conclude that it suffices to establish mirror symmetry between two particular
Calabi-Yau spaces $K$ and $\tilde{K}$ to establish it in the whole parameter
space. It would therefore be very interesting to find an example where mirror
symmetry may be rigorously proven.

Before giving such an example, we must briefly discuss the so called $A_k$
minimal models, which in a sense are the simplest examples of $N = 2$
superconformal field theories \cite{Capelli-Itzykson-Zuber}. The $A_k$ minimal
model has a Lagrangian representation as a Landau-Ginzburg theory with a single
field $X$ and the potential $X^{k+2}$ \cite{Witten-2}, but it may also be
defined in an abstract algebraic way in terms of its Hilbert space and algebra
of observables. It is essentially exactly solvable by the representation theory
of the $N = 2$ superconformal algebra. The following facts are of particular
importance to us \cite{Gepner}\cite{Gepner-Qiu}:

\vspace*{3mm} \begin{center} \parbox{127mm}
{\it The $A_k$ minimal model is invariant under a symmetry group isomorphic to
$\Z_{k+2}$. Furthermore, the orbifolds of the model with respect to $\Z_m$ anc
$\Z_{\tilde{m}}$subgroups of this group are isomorphic if $m \tilde{m} = k+ 2$
.}
\end{center}

We may now go back to the problem of finding an example of a mirror pair of
string compactifications. Consider the Landau-Ginzburg model with potential
\be
W(X_1, \ldots, X_5) = X_1^5 + \ldots + X_5^5. \label{Fermat}
\ee
We will refer to this particular model as the `Fermat point'. It is invariant
under $\Z_5^5$ acting as
\be
(X_1, \ldots, X_5) \rightarrow (\alpha_1 X_1, \ldots,  \alpha_5 X_5),
\ee
where the $\alpha_i$ are possibly different fifth roots of unity, \ie
$\alpha_1^5 = \ldots = \alpha_5^5 = 1$. We may thus construct Landau-Ginzburg
orbifolds $W/S$ for any subgroup $S$ of $\Z_5^5$. As before, the orbifold
$W/\Z_5$ for the `diagonal' embedding of $\Z_5$ in $\Z_5^5$ is the $r
\rightarrow - \infty$ limit of the corresponding Calabi-Yau theory. We see that
the Landau-Ginzburg model with the potential (\ref{Fermat}) is `separable' into
five $A_3$ minimal models, each described by a single field $X$ and the
potential $X^5$. Using this relationship and the previous result about the
orbifolds of the minimal models, it is possible to prove that
\be
W/\Z_5 \simeq (W/\Z_5)/\Z_5^3 = W/\Z_5^4
\ee
for a certain embedding of  $\Z_5^3$ and $\Z_5^4$ in $\Z_5^5$
\cite{Greene-Plesser}. This isomorphism is an example of mirror symmetry. From
our previous arguments, mirror symmetry now follows for other values of the
`shape' and `size' parameters in the whole parameter space of quintic
Calabi-Yau and Landau-Ginzburg models.

\vspace*{10mm} \noindent

{\bf 2. Mirror symmetry for the Kazama-Suzuki models}\\

We have seen that a crucial ingredient in the proof of mirror symmetry for the
quintic Calabi-Yau model and its Landau-Ginzburg relatives is the pairwise
equivalence of the orbifolds of the $A_k$ minimal models. As already mentioned,
this equivalence may be rigorously established by algebraic means using the
abstract definition of these models. However, it would be interesting to
understand the mirror symmetry property of the minimal models  in a Lagrangian
framework, where the models are defined by an action functional of some fields
and a path-integral over these fields. We have already given an example of such
a Lagrangian realization of the $A_k$ minimal model, namely a Landau-Ginzburg
theory with a single field $X$ and the potential $X^{k+2}$.

In this particular case, there is also a well supported conjecture for the
correct kinetic energy term \cite{Grisaru-Zanon}. However, the Landau-Ginzburg
theory is a strongly interacting quantum field theory and quite difficult to
work with. To find another Lagrangian representation, more suited to our needs,
we will first generalize the minimal models to the so called Kazama-Suzuki
models \cite{Kazama-Suzuki}. Such a model is specified by a Lie group $G$ with
a subgroup $H$ of the form $H \simeq U(1) \times H^\prime$ and a positive
integer $k$. The corresponding $N = 2$ superconformal field theory is referred
to as the $G/H$ Kazama-Suzuki model at level $k$.

It should be noted that the construction only works for certain groups $G$ and
$H$, though. The $A_k$ minimal model is equivalent to the $SU(2)/U(1)$
Kazama-Suzuki model at level $k$. A general Kazama-Suzuki model has a
Lagrangian representation which could be thought of as describing a string
propagating on the group manifold of $G$ such that the theory has a gauge
symmetry with the gauge group isomorphic to $H$
\cite{Schnitzer}\cite{Witten-3}.

We will now sketch how mirror symmetry for the Kazama-Suzuki models may be
understood using this Lagrangian representation. The  detailed calculation is
presented in \cite{Henningson}. The formulation of mirror symmetry for the
$A_k$ minimal model involves a symmetry group isomorphic to $\Z_{k+2}$. Our
first task is therefore to understand how a discrete symmetry group arises in
the Lagrangian formulation of  the $G/H$ Kazama-Suzuki model at level $k$.
Locally in field space, we may describe the model in terms of a scalar field
$\phi$ which is periodic with period $2 \pi$, a gauge field $A$ for the
$U(1)$-part of the gauge group $H \simeq U(1) \times H^\prime$, and some other
fields which need not concern us here. We now consider transformations of the
form
\be
\phi \rightarrow \phi + \gamma,
\ee
where the parameter $\gamma$ is a constant. The effective action $S$ of the
theory is not invariant under this transformation, but transforms as
\be
S \rightarrow S + \gamma (k + Q) \frac{1}{2 \pi} \int d A. \label{S-transf}
\ee
Here $Q$ is an integer called the dual Coxeter number of the group $G$. For the
case of  $G \simeq SU(2)$, which is relevant for the minimal models, we have $Q
= 2$. The integral in (\ref{S-transf}) is over the world-sheet. Naively, the
integrand is a total derivative, so since the world-sheet has no boundary one
would expect the integral to vanish by Stokes' theorem. However, the gauge
field $A$ need not be globally well-defined over the world-sheet, but may be
given by different one-forms over different patches related by gauge
transformations on the overlaps. The integrand is in fact a representative of
the first Chern class of the line bundle on which $A$ is a connection, and from
the theory of characteristic classes it follows that the integral,  including
the prefactor $(2 \pi)^{-1}$, may take any integer value.

Let us now take the parameter $\gamma$ to be a multiple of $2 \pi (k +
Q)^{-1}$. From the above discussion, it follows that the effective action $S$
then changes by a multiple of $2 \pi$. In a path integral, the action only
appears as $\exp i S$, and this quantity is thus invariant under the
transformation. Since $\gamma = 2 \pi$ acts trivially (because of the $2 \pi$
periodicity of $\phi$), we have identified a discrete symmetry group isomorphic
to $\Z_{k+Q}$. For the case of the $A_k$ minimal models, this is the $\Z_{k+2}$
symmetry that we discussed in the previous section.

Let us now consider a $\Z_m$ orbifold of the $G/H$ Kazama-Suzuki model at level
$k$. According to our general discussion of orbifolds, we may construct the
$\Z_m$ orbifold by changing the periodicity of $\phi$ from $2 \pi$ to $2 \pi /
m$. The theory is given by an action functional of the form
\be
S = \frac{k+Q}{2 \pi} \int d^2 z \, (\partial_z \phi \partial_{\bar{z}} \phi +
B_z \partial_{\bar{z}} \phi + B_{\bar{z}} \partial_z \phi ) + C, \label{action}
\ee
where $z$ and $\bar{z}$ are local coordinates on the world-sheet and $B_z$,
$B_{\bar{z}}$ and $C$ are some expressions independent of $\phi$. The important
point about this action is that the field $\phi$ only appears as $\partial_z
\phi$ and $\partial_{\bar{z}} \phi$. We may therefore write an equivalent first
order action by  introducing a vector field $V$ and replacing $\partial_z \phi$
and $\partial_{\bar{z}} \phi$ by $V_z$ and $V_{\bar{z}}$ respectively. The
relationship between $V$ and $\phi$ may be enforced by introducing a Lagrange
multiplier $\tilde{\phi}$, which we take to be periodic with period $2 \pi /
\tilde{m}$. Here $m$ and $\tilde{m}$ are related through $m \tilde{m} = k + Q$.
For $\Z_m$ to be a subgroup of $\Z_{k + Q}$, $m$ must divide $k + Q$ so that
$\tilde{m}$ is an integer. The first order action is
\be
S_1 = \frac{k+Q}{2 \pi} \int d^2 z \, (V_z V_{\bar{z}} + B_z V_{\bar{z}} +
B_{\bar{z}} V_z ) + C + \frac{k+Q}{2 \pi} \int d^2 z \, \tilde{\phi}
(\partial_z V_{\bar{z}} - \partial_{\bar{z}} V_z). \label{action1}
\ee
By performing the path integral over $\tilde{\phi}$ in this action we retrieve
the original action (\ref{action}). Indeed, $\tilde{\phi}$ acts as a Lagrange
multiplier for the constraint $d V = 0$, which we may solve locally, by the
Poincar\'e lemma, as $V = d \phi$. Furthermore,  since $\tilde{\phi}$ is $2 \pi
/ \tilde{m}$ periodic, it may have non-trivial winding around non-contractible
closed curves on the world-sheet. The path integral over $\tilde{\phi}$
contains a sum over the winding number  around each such curve, and this sum
constrains the holonomy of $V$ around the curve, \ie $\oint V$, to be a
multiple of $2 \pi / m$ \cite{Rocek-Verlinde}. This amounts to the field $\phi$
being $2 \pi /m$ periodic, so the action (\ref{action1}) indeed describes the
$\Z_m$ orbifold.

Since the first order action (\ref{action1}) is at most bilinear in the vector
field $V$ and the coefficient before the bilinear term is independent of the
other fields, the path integral over $V$ may be performed exactly. The result
is a new action $S_{\rm dual}$. A detailed computation yields that $S_{\rm
dual}$ equals the original action $S$ in (\ref{action}) with the field $\phi$
replaced by $\tilde{\phi}$. Since the latter field is $2 \pi / \tilde{m}$
periodic, this describes the $\Z_{\tilde{m}}$ orbifold. We have thus shown that
the $\Z_m$ and the $\Z_{\tilde{m}}$ orbifolds of the $G/H$ Kazama-Suzuki model
at level $k$ are equivalent if $m \tilde{m} = k + Q$, where $Q$ is the dual
Coxeter number of $G$.

In the above discussion we have omitted an important point: There exist two in
general inequivalent versions of the $G/H$ Kazama-Suzuki model at level $k$.
They are usually referred to as the vectorially and the axially gauged model.
The mirror transformation described above not only takes a $\Z_m$ orbifold into
a $\Z_{\tilde{m}}$ orbifold, but also interchanges the vector model and the
axial model. However,  for the case of $G/H \simeq SU(2)/U(1)$ the two models
are related by a change of variables under which the measure in the
path-integral is invariant, so they both describe the same $N = 2$
superconformal field theory, \ie the $A_k$ minimal model. We have thus proven
the mirror symmetry property of the $A_k$ minimal models.

\vspace*{10mm} \noindent
{\bf 3. Conjugate Landau-Ginzburg models}\\

The Fermat point Landau-Ginzburg model with potential
\be
W(X_1, \ldots, X_5) = X_1^5 + \ldots + X_5^5
\ee
is conjugate to itself in the sense that the orbifolds $W/S$ and $W/\tilde{S}$
are equivalent when $S$ and $\tilde{S}$ are `dual' subgroups of the symmetry
group $\Z_5^5$. We have already discussed a particular example of such
equivalences when we constructed the mirror partner of (the $r \rightarrow -
\infty$ limit of) the quintic Calabi-Yau model.

It is now natural to consider more general Landau-Ginzburg models described by
a set of fields $X_1, \ldots, X_n$ and a (quasi)homogeneous potential $W(X_1,
\ldots, X_n)$. This means that if we assign a degree to each of the fields
$X_1, \ldots, X_n$ then all the terms in the potential $W(X_1, \ldots, X_n)$
should be of the same degree. We then look for pairs of potentials $W(X_1,
\ldots, X_n)$ and $\tilde{W}(\tilde{X}_1, \ldots, \tilde{X}_n)$ with isomorphic
symmetry groups such that the orbifolds $W/S$ and $\tilde{W}/\tilde{S}$ are
equivalent when $S$ and $\tilde{S}$ are `dual' subgroups of the symmetry group.

To implement this program, we need a way to compare two Landau-Ginzburg
orbifolds and determine if they might be equivalent. For example, one might try
to compare the Hilbert spaces of the corresponding $N = 2$ superconformal field
theories. Each state in the Hilbert space is characterized by its `energy' and
some other quantum numbers. If the two theories have identical spectra of all
quantum numbers, it would be a strong indication that they are in fact
equivalent.

Unfortunately, it is in general not possible to calculate the spectrum of a
Landau-Ginzburg theory. One difficulty is that the action of the theory is not
known, since the exact form of the kinetic energy is in general unknown.
Furthermore, the Landau-Ginzburg theories are strongly interacting quantum
field theories, and therefore difficult to analyze. We must therefore content
ourselves with less complete information than the complete spectrum. A
tractable calculation is to restrict our attention to the spectrum of
zero-energy states and count bosonic and fermionic states with opposite sign.
This is usually referred to as calculating the `Poincar\'e polynomial'
\cite{Intriligator-Vafa} or the `elliptic genus'
\cite{Witten-2}\cite{DiFrancesco-Yankielowicz} of the theory.

The reason that these quantities are effectively calculable may be found by
considering the supersymmetry algebra \cite{Witten-5}. Schematically, this
algebra can be thought of as being generated by a bosonic energy generator $H$
and a fermionic supersymmetry generator $Q$. The generators $H$ and $Q$
commute, and the square of $Q$ equals $H$. Since $H$ may be written as the
square of a self-adjoint operator, we may conclude that all energy eigenvalues
are non-negative. Suppose now that we have a bosonic state $|{\rm bos}>$ of
strictly positive energy, \ie $H |{\rm bos}> = h |{\rm bos}>$ for some $h > 0$.
We may then define a fermionic state $|{\rm ferm}>$ by acting on $|{\rm bos}>$
with the supersymmetry generator, \ie $|{\rm ferm}> = Q |{\rm bos}>$.

{}From the properties of the supersymmetry algebra, it easily follows that
$|{\rm
ferm}>$ has the same energy eigenvalue as $|{\rm bos}>$, \ie $H |{\rm ferm}> =
h | ferm>$. States of non-zero energy thus appear in Bose-Fermi pairs with the
same values of all other quantum numbers which commute with $Q$. The argument
fails for zero-energy states, because $Q$ annihilates such a state rather than
creating a new state of opposite statistics. We may therefore have an unequal
number of bosonic and fermionic zero-energy states for a given set of values of
the remaining quantum numbers. If we now deform the theory smoothly, the energy
eigenvalues will change continuously.

It might then happen a Bose-Fermi pair of non-zero energy descends down to zero
energy. However, this does not change the number of bosonic zero-energy states
minus the number of fermionic zero-energy states for given values of the other
quantum numbers. This number may therefore be calculated by deforming the model
to some well understood theory, typically by turning of all interactions so
that we get a free field theory. We may then search for possible mirror pairs
by comparing the zero-energy spectra of two theories in this way. Obviously,
this rather crude method is not sensitive to continuous deformations such as
the `shape' and `size' deformations.

We will now briefly present the main results of this investigation. For a more
detailed review see \cite{Berglund-Henningson-II} or the original paper
\cite{Berglund-Henningson-I}. Our main result is that we found two classes of
possible pairs of conjugate Landau-Ginzburg potentials.  (These models were
first proposed in \cite{Berglund-Hubsch}.) The first class of examples relates
the potentials
\begin{eqnarray}
W(X_1, \ldots, X_n) & = & X_1^{\alpha_1} + X_1 X_2^{\alpha_2} + \ldots +
X_{n-1} X_n^{\alpha_n} \nonumber\\
\tilde{W}(\tilde{X}_1, \ldots, \tilde{X}_n) & = & \tilde{X}_1^{\alpha_n} +
\tilde{X}_1 \tilde{X}_2^{\alpha_{n-1}} + \ldots + \tilde{X}_{n-1}
\tilde{X}_n^{\alpha_1} .
\end{eqnarray}
Both of these potentials are invariant under a symmetry group isomorphic to
$\Z_D$ for $D = \alpha_1 \ldots \alpha_n$. A comparison of the spectra of
zero-energy states shows that suitable deformations of the Landau-Ginzburg
orbifolds $W/\Z_m$ and $\tilde{W}/\Z_{\tilde{m}}$ might be each others mirror
partners when $m \tilde{m} = D$. We note that the case $n = 1$ gives the $A_k$
minimal models, for which mirror symmetry has been rigorously established by
algebraic methods and by the method described in the previous section.

Our second class of examples relates the potentials
\begin{eqnarray}
W(X_1, \ldots, X_n) & = & X_n X_1^{\alpha_1} + X_1 X_2^{\alpha_2} + \ldots +
X_{n-1} X_n^{\alpha_n} \nonumber\\
\tilde{W}(\tilde{X}_1, \ldots, \tilde{X}_n) & = & \tilde{X}_n
\tilde{X}_1^{\alpha_n} + \tilde{X}_1 \tilde{X}_2^{\alpha_{n-1}} + \ldots +
\tilde{X}_{n-1} \tilde{X}_n^{\alpha_1} .
\end{eqnarray}
The symmetry groups are isomorphic to $\Z_D$ for $D = \alpha_1 \ldots \alpha_n
+ (-1)^{n-1}$, and the zero-energy spectra indicate a possible mirror symmetry
between deformations of the Landau-Ginzburg orbifolds $W/\Z_m$ and
$\tilde{W}/\Z_{\tilde{m}}$ when $m \tilde{m} = D$. Again, the $n = 1$ case
gives the $A_k$ minimal models.

Finally, one may take any combination of these examples. The total symmetry
group is then the product of the symmetry groups of the simple theories, and
the zero-energy spectra of the Landau-Ginzburg orbifolds with respect to `dual'
subgroups of the total symmetry group are consistent with mirror symmetry. We
have in fact already given an example of this procedure when we regarded the
quintic Landau-Ginzburg theory at the Fermat point as a product of five $A_3$
minimal models.

\vspace*{5mm}This work is supported by DOE grant DE-FG02-92ER40704.

\end{document}